\documentclass[a4]{elsart}

\usepackage{isolatin1}
\usepackage{graphicx}
\usepackage{color}
\usepackage{amssymb}
\usepackage{amsmath}
\usepackage{latexsym}
\newcommand{\dfracd}[2]{\dfrac{\text{d} #1}{\text{d} #2}}

\DeclareGraphicsExtensions{.eps}
\begin{document}

\begin{frontmatter}

\title{Mathematical Model of HIV superinfection dynamics and R5 to X4 switch}

\author{L. Sguanci\thanksref{a}\thanksref{c}},
\author{P. Li\`{o}\thanksref{b}\thanksref{1}},
\author{F. Bagnoli\thanksref{a}\thanksref{c}}

\address[a]{Dipartimento di Energetica, Universit\`{a} di Firenze, Via S. Marta 3, 50139 Firenze, Italy.}
\address[b]{Computer Laboratory, University of Cambridge, CB3 0FD Cambridge, UK.}
\address[c]{INFN, sezione di Firenze and CSDC.}

\thanks[1]{pietro.lio@cl.cam.ac.uk}

\begin{abstract}
During the HIV infection several quasispecies of the virus arise,
which are able to use different coreceptors, in particular the
CCR5 and CXCR4 coreceptors (R5 and X4 phenotypes, respectively).
The switch in coreceptor usage has been correlated with a faster
progression of the disease to the AIDS phase. As several
pharmaceutical companies are starting large phase III trials for
R5 and X4 drugs, models are needed to predict the co-evolutionary
and competitive dynamics of virus strains. We present a  model of
HIV early infection which describes the dynamics of R5 quasispecies
and a model of HIV late infection which describes the R5 to X4
switch. We report the following findings: after superinfection or
coinfection, quasispecies dynamics has time scales of several
months and becomes even slower at low number of CD4+ T cells. The
curve of CD4+ T cells decreases, during AIDS late stage, and can
be described taking into account the X4 related Tumor Necrosis
Factor dynamics. Phylogenetic inference of chemokine receptors
suggests that viral mutational pathway may generate R5 variants
able to interact with chemokine receptors different from CXCR4.
This may explain the massive signaling disruptions in the immune
system observed during AIDS late stages and may have relevance for
vaccination and therapy.
\end{abstract}

\end{frontmatter}

\section{Introduction}
\label{sec:Introduction}

Human immunodeficiency virus type 1 (HIV-1) infection is
characterized by the progressive loss of CD4+ T cells. Infection
by most strains of HIV-1 requires interaction with CD4 and a
chemokine receptor, either CXCR4 or CCR5. During early stages of
HIV-1 infection, viral isolates most often use CCR5 to enter cells
and are known as R5 HIV-1. Later in the course of HIV-1 infection,
viruses that use CXCR4 in addition to CCR5 (R5X4) or CXCR4 alone
(X4 variants) emerge in about 50\% patients (switch virus
patients)~\cite{KF2004,GG2005}. These strains are
syncytium-inducing and are capable of infecting not only memory T
lymphocytes but also naïve CD4+ T cells and thymocytes through the
CXCR4 coreceptor. The switch to use of CXCR4 has been linked to an
increased virulence and with progression to AIDS, probably through
the formation of cell syncytia and killing of T cell precursors.
X4 HIV strains are rarely, if ever, transmitted, even when the
donor predominantly carries X4 virus. CXCR4 is expressed on a
majority of CD4+ T cells and thymocytes, whereas only about 5 to
25\% of mature T cells and 1 to 5\% of thymocytes express
detectable levels of CCR5 on the cell surface~\cite{GS2005}. It is
noteworthy that X4 HIV strains stimulate the production of
cellular factor called Tumor Necrosis Factor (TNF), which is
associated with immune hyperstimulation, a state often implicated
in T-cell depletion~\cite{HS2005a}. TNF seems able to both inhibit
the replication of R5 HIV strains while having no effect on X4 HIV
and to down regulate the number of CCR5 co-receptors that appear
on the surface of T-cells~\cite{PY2004}.

A powerful concept in understanding HIV variability and its
consequences is that of quasispecies~\cite{ES1989}. Quasispecies
are the combined result of mutations and recombination, that
originate variability, and of  co-infection (simultaneous
infection), superinfection (delayed secondary infection) and
selection, that keep variability low. HIV-1-infected individuals
show heterogeneous viral populations of related genomes best
described as viral quasispecies~\cite{BE2005}.
Infact, the infection capacity of mutants may vary, and also their
speed of replication~\cite{NP1998}. Moreover, since the number of
targets (the substrate) is limited, fitter clones tend to
eliminate less fit mutants, which are subsequently regenerated by
the mutation mechanism~\cite{ES1977}. While mutations are an
essential ingredient for exploring the genetic space in the search
for the fitness maximum, they also lowers the average fitness of
the strain, that generally is formed by a cloud of mutants around
the fitness maximum, the quasispecies. It's worth noting that, for
a given fitness landscape, there is a maximum tolerable mutation
rate above which the quasispecies structure is lost (error
threshold~\cite{ES1989}).

The use of mathematical models is an insightful and essential complement to in vivo and in vitro experimental design and interpretation. Indeed mathematical models of HIV dynamics have proven valuable in understanding the mechanisms of many of the observed features of the progression of the HIV infection~\cite{HM1995,CP2004,CS1996,DP1995,WN2002,PH1996,WP2004,WS1995}. With the incorporation of accurate stoichiometries, gene expression levels and detailed kinetic information, from bioinformatics of sequence analysis and molecular dynamics, mathematical models will be even more effective in predicting time and space patterns and the effects of drug therapy.

Here we address the issue of studying the coevolutive and competitive dynamics of different strains of HIV-1 virus also leading to the R5 to X4 phenotype switching. In doing this in the next section we introduce two models: a quasispecies model for R5 phase in which several R5 strains appear by mutations, co-infection and superinfection. In the limit of 1 quasispecies we are able to find the same values observed experimentally and in other models (most notably Perelson's standard model). We test the model in the scenarios of co-infection and superinfection using parameters derived from biological literature. The second model focuses on the R5 to X4 shift and the hyperstimulation of T cell precursors through TNF. We are able to describe the decreasing dynamics of CD4+ after the appearance of X4 strains and make predictions on the HAART results in coinfection and superinfection scenarios at different times of the disease progression. We make use of phylogenetic models of the amino acid sequences of the human and mice chemokine receptor families to investigate the mutational pathway underlying the switching from different chemokine receptors. Finally we find that the switch from R5 to X4 may allow the HIV to bind to other chemokine receptors.

\section{Models}

\subsection{A model of the early R5 phase}

In someone who is newly infected by HIV, several variants of the
virus, called R5, are often the only kind of virus that can be
found. A meaningful way to model strain mutations, coinfection and
superinfection is to extend Perelson's standard
model~\cite{PH1996} to multiple strains and incorporate immune
response to R5 quasi species amplification (in the following termed QSR5 model). The R5  quasispecies
dynamics can be described by the following set of equations:

\begin{align}
\dot{T}_{i}  &= \left(
\lambda_{i}+\sum_{k}\gamma_{ik}^{(T)}I_{k}T_{i}\right)
  \left(  1-\frac{1}{K}\sum_{i}T_{i}\right)  -
  \left(  \delta_{T}+\sum_{k}\beta_{k}V_{k}\right)T_{i}  ,\label{Ti}\\
\dot{I}_{k}  &= \left(  \sum_{k^{\prime}}\mu_{kk^{\prime}}
\beta_{k^{\prime}}V_{k^{\prime}}\right)\left(\sum_{i}T_{i}\right)
 -\left(  \delta_{I}+\sum_{i}\gamma_{ki}^{(I)}T_{i}\right)I_{k}
 ,\label{Ik}\\
\dot{V}_{k}  &= \pi
I_{k}-\left(c+\sum_{i}\gamma_{ki}^{(V)}T_{i}\right)V_{k}.\label{Vk}
\end{align}

The following cell types are considered: T-helper (CD4+) cells
carrying the CCR5 co-receptor responding to virus strain $i$,
($T_{i}$); T cells infected by virus strain $k$, ($I_{k}$); $k$
strains of R5 virus, ($V_{k}$). We have thus assumed that viral
strain $k$ are identified by just one epitope, which is then
displayed on the surface of the T cell of class $k$,  and that a T
cell of class $i$ can be activated at least by one CD4+ T cell
carrying the epitope $k$, which is specific of the viral strain
$k$. The indices $i$ ($k$) range from 1 to $N_i$ ($N_k$), and in
the following we have used $N_i=N_k=N$.

In Equation~\eqref{Ti} T cells are generated through two
mechanisms: the bone-marrow source (and selection in the thymus)
and the duplication of T cell strains activated upon the
recognition with an antigen carrying cell that may be even an
infected one. We modeled T cells activation as a logistic term
mimicking the global carrying capacity of immune
system~\cite{BS2005}.

The death-rate term is composed by a natural death rate
proportional to the population, and by the infection rate of T
cells due to any viral strain. The term
$\sum_{k}\beta_{k}V_{k}T_{i}$ and the sum over $T_{i}$ in the $I$
cell birth rate reproduce the infection probability, that is the
same irrespective of the T class. As a cell become infected it
does no more contribute to the immune response.

Equation~\eqref{Ik} describes the infection dynamics. The two
death rate parameters account for the decrease of the infected
cells due to cellular death and after the action of T-killer cells
(CD8+). Even if there are clear experimental evidences that CD4+
cells decrease during the late HIV infection stages and in the
AIDS state, as far as the asymptomatic phase of the infection is
concerned, the parameter $\delta^I$ may be assumed as a constant,
medical literature referred, value.

The $\mu_{kk^{\prime}}$ term is responsible of the mutation
process affecting the phenotype, essential for the formation of
new quasispecies. The choice of a mutation rate of the order of
$10^{-5}$ is based on considering only those non-synonymous
mutations that alter the phenotype (protein
structure)~\cite{BS2005}.

In Equation~\eqref{Vk} the virus replicative dynamics is
described. The birth rate term is proportional to the virus
"budding" numerosity while the viral death rate parameters depend
on the  rate of natural death and accounts for the recognition of
virus by B cells.

It's worth noting that B-cells and T-killer cells are only
implicitly included in the model in order to reduce the
dimensionality without loosing too many details. We assume that
these responses are fast enough to be at equilibrium and they are
just proportional to the abundance of (cognate) $T$ helper cells.

The three $\gamma$ parameters $\gamma^{(T)}$,
$\gamma^{(I)}$ and $\gamma^{(V)}$ are matrices
describing the interactions between cells and/or cells and
viruses, i.e.\ who will interact with whom, in terms of geometry
and strength of the interaction. It is thus possible to consider
which strains of the virus are recognized and with which accuracy,
and the same for the action of T-killer cells and B-cells. It is
also worth noting that $\gamma^{(T)}_{ii}$ is the most important
determinant of the viral fitness~\cite{BS2005}.

\subsection{Modeling the transition R5 to X4}
In about half of the people who develop advanced HIV disease, the
virus begins to use another co-receptor called CXCR4 (X4 viral phenotype). The
shift to using CXCR4 is considered a bad sign because it is often
accompanied by a dramatic increase in the rate of T-cell
depletion. The inability of the thymus to efficiently compensate
for even a relatively small loss of naïve T cells may be a key
factor for CD4+ T cells depletion and AIDS progression. We
hypothesize that it may not be exhaustion of homeostatic
responses, but rather thymic homeostatic inability along with
gradual wasting of T cell supplies through hyperactivation of the
immune system that lead to CD4 depletion in HIV-1 infection.
\begin{figure}[t]
  \centering
  \includegraphics[height=0.5\textheight]{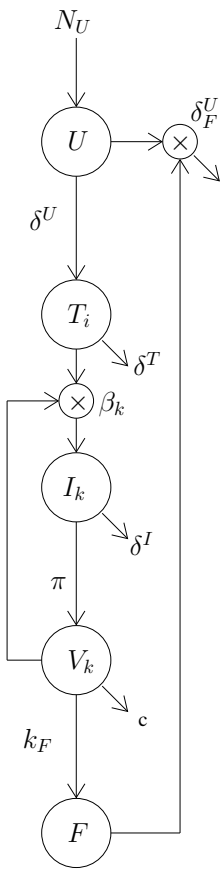}
  \caption{Schematic description of the model for the switching from R5 to X4 viral phenotype. Na\"ive T-cells, $U$, are generated at constant rate $N_U$ and removed at rate $\delta^U$. They give birth to differentiated, uninfected T-cells, $T$. These in turn are removed at constant rate $\delta^T$ and become infected as they interact with the virus. Infected T-cells, $I$, die at rate $\delta^I$ and contribute to the budding of viral particles, $V$, that are cleared out at rate $c$. As soon as the X4 phenotype arise, the production of the TNF starts, proportional to the X4 concentration and contribute to the clearance of na\"ive T-cells, via the $\delta^{U}_{F}$ parameter.}
  \label{fig:modelB}
\end{figure}
We here introduce a modified version of the previous model, that
considers only CD4 dynamics. Neither B nor CD8+ T cells are
explicitly modeled and the space that now is considered is that of
different phenotypes of the virus (we term this model R5toX4 model). Under those assumptions we may
focus on the  appearance of X4 viruses and on their subsequent
interaction with R5 strains.

\begin{eqnarray}
\dfracd{U}{t} &=& N_{U} - \delta^{U} U -
\delta^{U}_{F} U F
\label{eq:U} \\
\dfracd{T_i}{t} &=& \delta^{U} U - \left( \sum_k \beta_k V_k
\right) T_i - \delta^T T_i
\label{eq:T} \\
\dfracd{I_k}{t} &=& \left(\sum_{k'} \mu_{kk'} \beta_{k'}
V_{k'}\right) \left(\sum_i T_i\right) - \delta^I I
\label{eq:I} \\
\dfracd{V_k}{t} &=& \pi I_k - c V_k
\label{eq:V} \\
\dfracd{F}{t} &=& k_{F} \sum_{k \in X4} V_k
\label{eq:TNF}
\end{eqnarray}

In the equations above, the variables modeled are the pool of
immature CD4+ T cells, $U$, the different strains of uninfected
and infected T cells ($T$ and $I$, respectively), HIV virus, $V$,
and the concentration of TNF, $F$. A schematic view of the model is depicted in
Fig.\ref{fig:modelB}. The value of the parameters introduced with respect to the R5
model are summarized in Table~\ref{table:param}.

{\centering
\begin{table}
{\scriptsize \textsf{
\begin{tabular}{llcc}
\hline
Parameter & Symbol & Value & Units of Meas.\\
\hline
Production of immature T cells & $N_U$ & 100 & cell/$\mu$l $t^{-1}$\\
Death rate of immature T cells & $\delta^U$ & 0.1 & $t^{-1}$\\
Death rate of immature T cells upon the interaction with TNF & $\delta^U_F$ & $10^{-5}$ & $\mu$l/cell $t^{-1}$\\
Decreasing infectivity of R5 phenotype due to TNF & $k_{R5}$ & $10^{-7}$ & $(\mu l/\text{cell})^2\ t^{-1}$\\
Increasing infectivity of R5 phenotype due to TNF & $k_{X4}$ & $10^{-7}$ & $(\mu l/\text{cell})^2\ t^{-1}$\\
Increasing death rate of immature T cells due to TNF & $\delta^I_{X4}$ & 0.0005 & $\mu$l/cell $t^{-1}$\\
Rate of production of TNF & $k_F$ & 0.0001 & $t^{-1}$\\
\hline
\end{tabular}
\\
\label{table:param}
\caption{Model for the R5 to X4 phenotypic switch: a summary of the additional parameters introduced. The value of the other parameters are medical literature referred, see also~\cite{BC2000}.}
}}
\end{table}
}

In particular, Equation~(\ref{eq:U}) describes the constant
production of immature T cells by the thymus $N_{U}$ and their
turning into mature T cells at rate $\delta^{U}$. If X4 viruses
are present, upon the interaction with TNF, immature T-cells are
cleared at fixed rate $\delta^{U}_{F}$.

Equation~(\ref{eq:T}) describes how uninfected mature T cells of strain $i$ are produced at fixed rate $\delta^{U}$ by the pool of immature T cells. Those cells, upon the interaction with any strain of the virus, $V_k$, become infected at rate $\beta_k=\beta \quad \forall k$. The infectiousness parameter, $\beta$, is not constant over time, but depends on the interplay between R5 and X4 viruses. In particular, due to the presence of TNF, the infectivity of R5 strains is reduced ($\beta_{R5}(t)=\beta-k_{R5}F(t)$), while the one of X4 viruses increases, with constant of proportionality $k_{X4}$ ($\beta_{X4}(t) = \beta + k_{X4}\ F(t)$), mimicking the cell syncytium effect induced by the TNF molecule.

Equation~(\ref{eq:I}) describes the infection of mature T-cells.
Infected T-cells of strain $k$ arise upon the interaction of a
virus of strain $k$ with any of the mature T-cell strains. The
infected cells, in turn, are cleared out at a rate $\delta^I$. When
TNF is released, this value increases linearly with constant
$\delta^I_{X4}$, $\delta^I(t) = \delta^I + \delta^I_{X4}\ F(t)$.

Equation~(\ref{eq:V}) is close to that in the R5 quasispecies model, a part from different viral phenotypes being here considered.

Finally, in Equation~(\ref{eq:TNF}), we model the dynamics of
accumulation of TNF by assuming the increase in TNF level to be
proportional, via the constant $k_{F}$, to the total
concentration of X4 viruses present.

\subsection{Investigating the mutational pathway from R5 to X4}
In our model we represent the different phenotypes by using a
linear strain space, see for instance~\cite{GG2002,LL2002} for
similar assumptions. The strain space is ordered in terms of
phenotype similarity. This assumption is justified if the
phenotypes are determined by few viral protein functional
determinants which are both independent and differ only few DNA
bases, i.e. few mutations can change one determinant into another.

Although we are aware of the several recent works on HIV
mutational dynamics and phylogenetic assessments, we thought that
a meaningful way to estimate the mutational pathways between R5
and X4 seen is to use phylogenetic inference on chemokine receptor
families. The assessment of phylogenies using likelihood framework
depends on the choice of an evolutionary model~\cite{WLG2001}. We computed the
maximum likelihood (ML) analysis of the CRs data set under
different models of evolution:~\cite{DO1978}, JTT~\cite{JT1994},
WAG~\cite{WG2001}. We used these models considering the
incorporation of the amino acid frequencies of the chemokine data
sets, (`+$F$'), and the heterogeneity of the rates of evolution,
implemented using a gamma distribution (`+$\Gamma $')~\cite{Y1994,Y1997}.

\section{Results}

We have first extended Perelson's standard model~\cite{PH1996}
to incorporate different antigen recognition abilities by the
immune system and coexistence dynamics of different R5 strains of
HIV virus. Our approach is a mean field one, i.e. we investigate the
average quantities of these molecular species~\cite{R2002,L2003,HH2004,AW2002,GH2005}. Then, we have
modified the model to describe features of the latent phase of the
infection, i.e. the R5 to X4 switch and the hyperstimulation of the T
cell precursors through the TNF.

\subsection{Amplification of R5 strains: mutation, co-infection and super-infection}
Recent works have shown that HIV quasispecies may compete~\cite{CD2000} and cooperate~\cite{VA2006} and that persistence of the initial or ancestor
quasispecies is a good indicator for disease progression
\cite{BL2005}. Burch and Chao~\cite{BC2000} have stressed that the
evolution of an RNA virus is determined by its mutational
neighborhood. As the phenotype divergence among viral strains
arises from differences in selection pressure, these differences
may lead, for instance, to a higher infection rate. Since the
competition is through the immune system response and given that
the phase space of antigen recognition is not homogeneously
covered~\cite{DP1992}, the HIV high mutation rate allows the
quasispecies to find regions with weak immune response. This
competition may lead to speciation of viral strains.

\begin{figure}[t]
  \centering
  \includegraphics[width=0.4\textwidth]{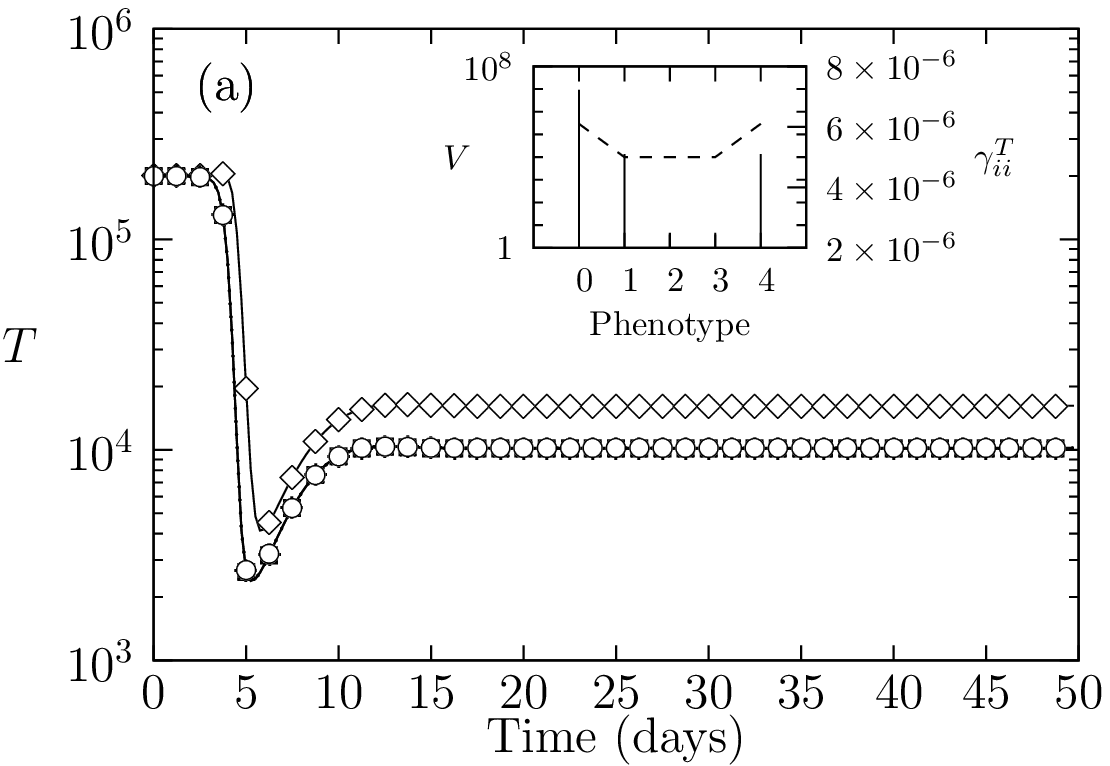}
  \includegraphics[width=0.4\textwidth]{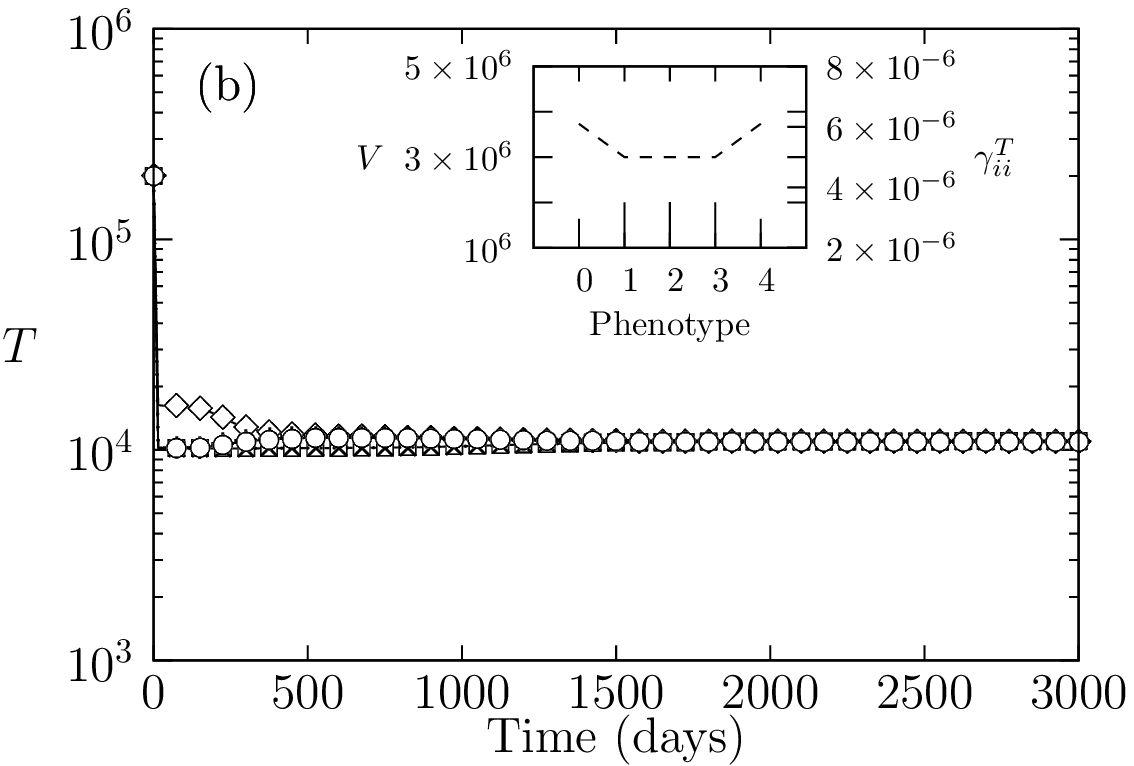}
  \caption{Typical time evolution of the T cell abundance
    in the R5 model for short-term (a) and long-term
    behavior (b). We set $\mu=10^{-5}$ with $N=5$.
    In the inset, the y-axis on the left reports the abundance of
    virus strains at time $t$=50 in plot (a) and $t$=3000 in plot (b);
        the y-axis on the right shows the interaction
    strength (dashed line) between T cells and virus phenotypes (x-axis).}
  \label{fig:ev}
\end{figure}

If we consider the model of the early phase of the infection, the evolution of T cells abundances in a scenario of quasispecies is shown in Fig.~\ref{fig:ev}a. As the asymptotic state of our model is a fixed point, the asymptotic distribution is insensitive of the initial conditions, and the strains corresponding to higher fitness are more abundant (see inset of Fig.~\ref{fig:ev}b). However, one should consider that this asymptotic state may be reached after such a long time that it may be outside any practical scenario of the progression of a disease. The role of mutations in the transitory regime is quite particular. First of all, starting from the first inoculum at time $t=0$ on the zeroth phenotype, mutations are necessary to populate the other strains of the virus, see also Fig.~\ref{fig:ev}. Moreover, in the presence of coupling among strains, due to competition or to a global constraints (the $K$ parameter), the specific form of mutations does not play a fundamental role, see also Ref.~\cite{BB1997}.

\begin{figure}[t]
  \centering
  \includegraphics[width=0.6\textwidth]{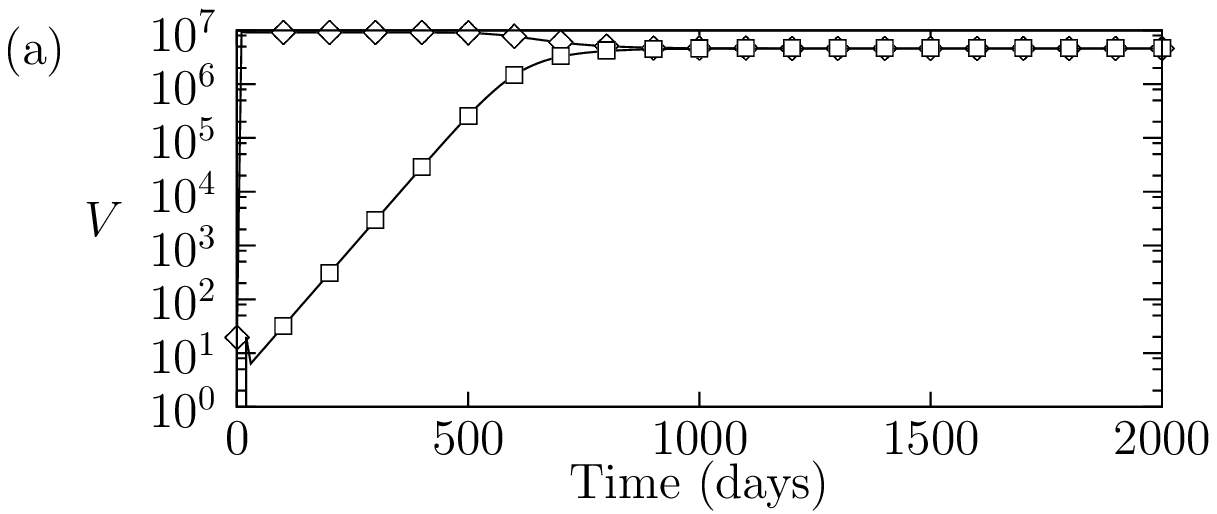}\\
  \includegraphics[width=0.6\textwidth]{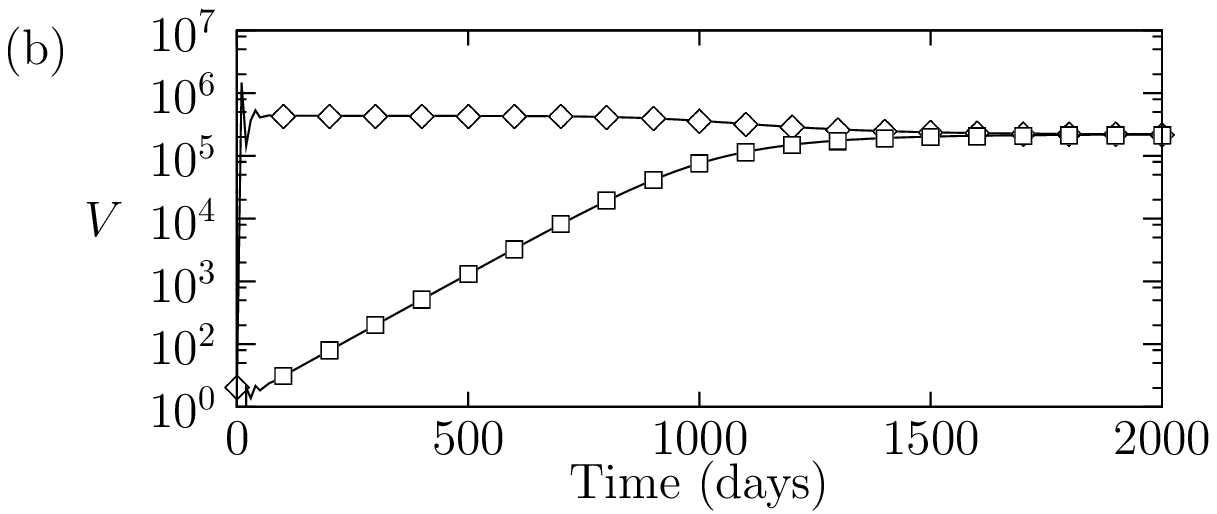}
  \caption{Viral counts, $V$, during a superinfection scenario. We set $N$=5 and no mutation is considered, thus $\mu$=0. (a) A slow mounting of the second viral infection ($\Box$), having time scale of several months, is observed. In (b) a compromised immune system is considered. The time for the second strain to reach the same abundance of the first-infecting strain ($\Diamond$) is greater than in (a).}
  \label{fig:super}
\end{figure}

Figs.~\ref{fig:super}a and~\ref{fig:super}b report the results of short time and long term viral coevolution after superinfection, respectively.
We considered the first viral inoculum to happen at time $t=0$, with the superinfection event occurring at time $t=20$, when the immune response to the first inoculum has completed and the virus has established a chronic infection.
After the second inoculum the model exhibits a short transient, followed by a slow mounting of the second infection. Due to the resulting low dynamics, the time needed by the second quasispecies to reach the same level of the other amounts to several months (Fig.~\ref{fig:super}a), and represents another example of a slow relaxation toward a fixed-point equilibrium.
We may also take into consideration the progression of the disease, characterized by a compromised immune system, assuming, as a first approximation, a lower number of T cells, \emph{i.e.} a lower value of $\lambda$, with respect to the previous scenario. In this case the strain corresponding to the second inoculum requires much longer time to reach the same abundances of the first strain (Fig.~\ref{fig:super}b).

If we consider HAART therapy, is interesting to address the question of what happens if a patient suddenly stops the drugs treatment. During HAART virus load in blood declines, but the virus is not definitively eradicated. At the same time T cells recover and their concentration arise towards the stationary state corresponding to the low-level concentration of the virus. As the treatment is interrupted, virus level begins to increase again. Something similar to a second, delayed infection, occurs. It's worth noting that, if more than one viral strain is present, as in the case of a coinfected or superinfected patient, the second infection may lead to a different asymptotic dominant strain. Infact the growth velocities of the different phenotypes of the virus depend both on the concentration of uninfected T cells and on the fitness of the different strains. Thus a more aggressive strain or one with a weaker recognition by the immune system may take advantage of the interruption of the drugs treatment and of the renewed infection, leading to an acceleration of the disease.

\begin{figure}[t]
  \centering
  \includegraphics[width=0.4\textwidth]{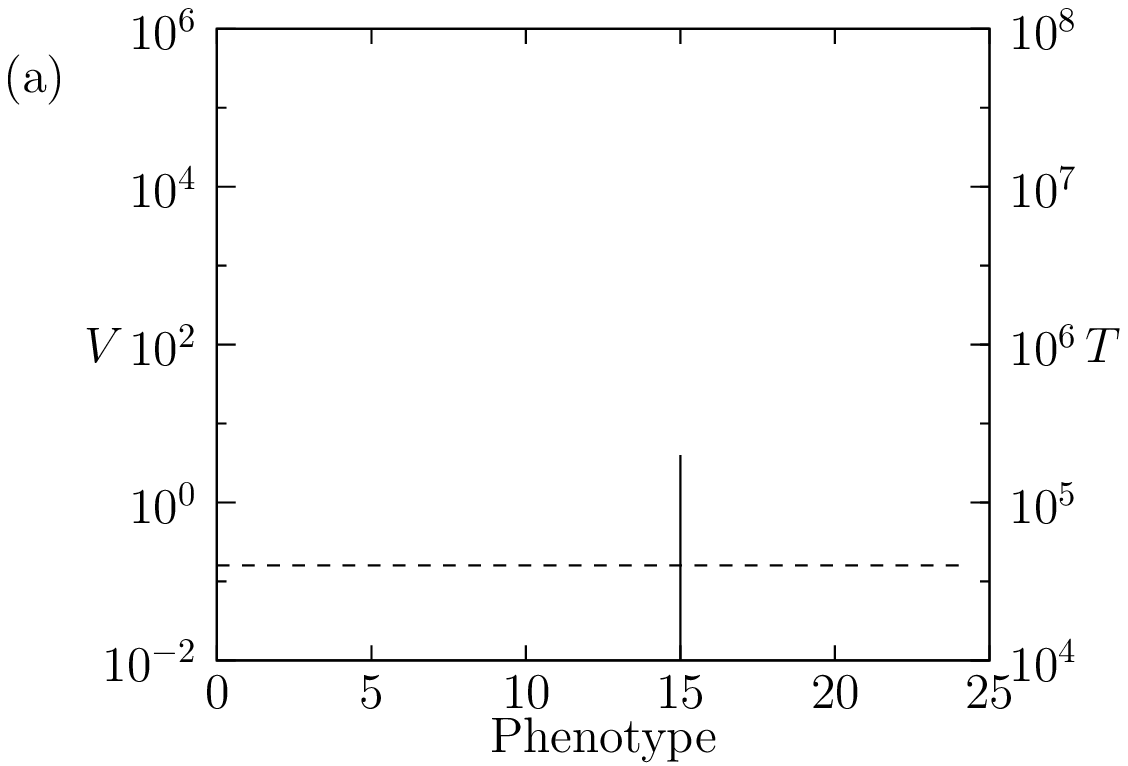}
    \includegraphics[width=0.4\textwidth]{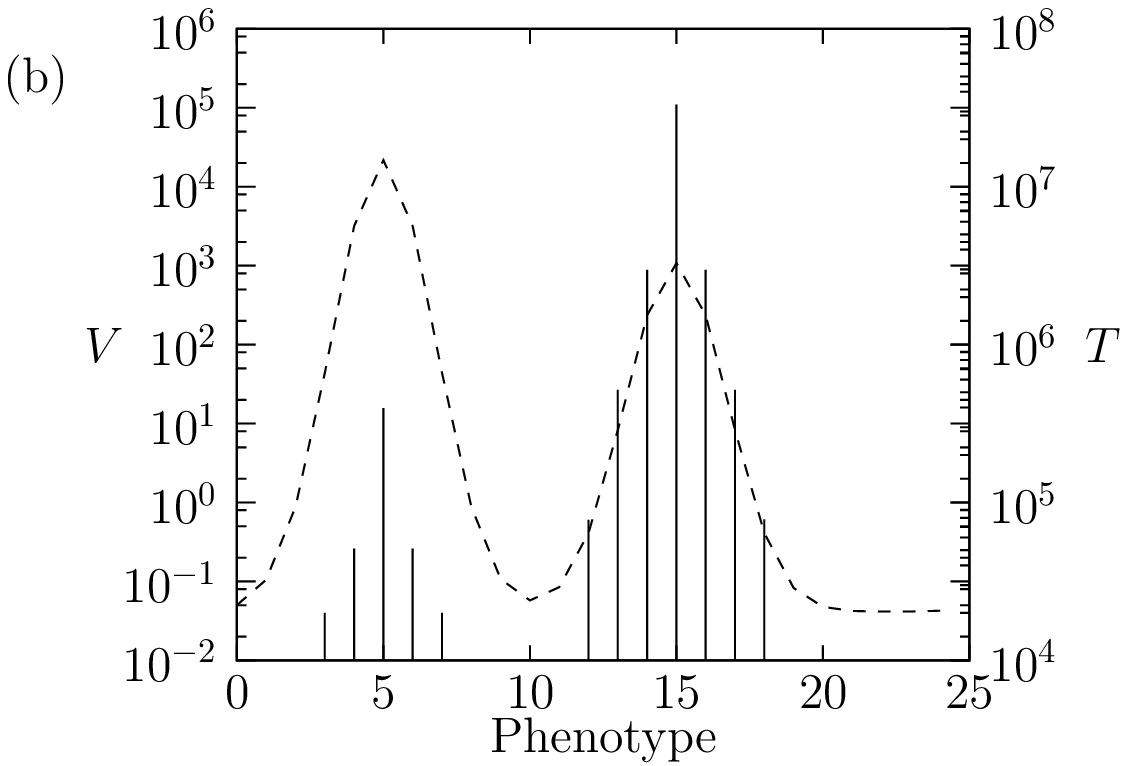} \\
  \includegraphics[width=0.4\textwidth]{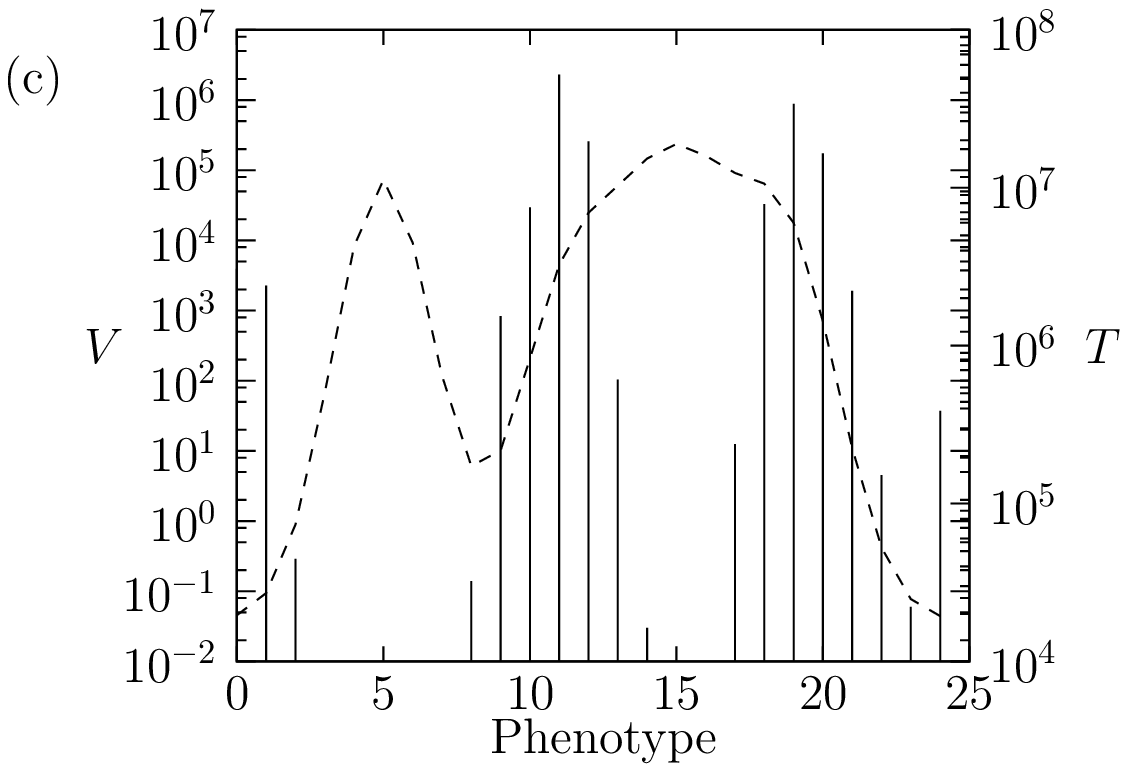}
    \includegraphics[width=0.4\textwidth]{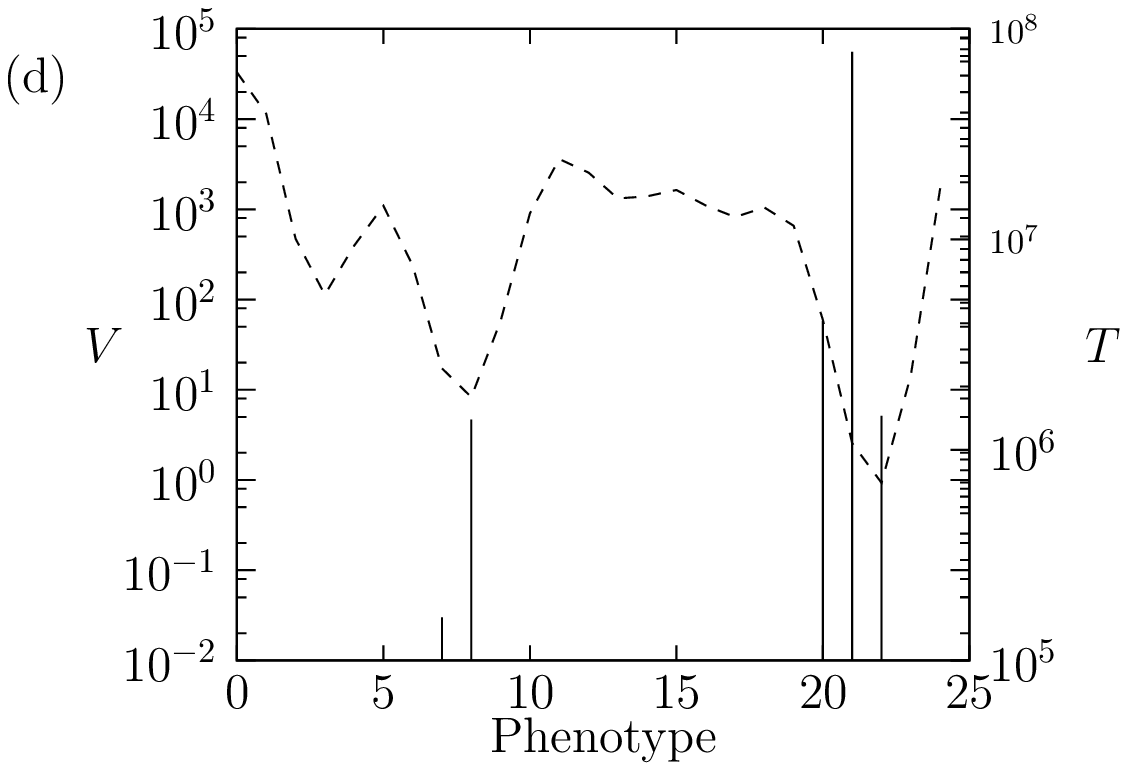} \\
  \caption{Speciation of virus quasispecies and uninfected T cells dynamics
after competitive superinfection at four different times: $t=0$
(a), $t=4.5$ (b), $t=5.25$ (c) and $t=5.75$ (d). Virus strain 15
is present at time $t=0$, while strain 5 is inoculated at time
$t=1$. Mutation rate $\mu=10^{-4}$ and non-uniform interaction
strength as in Fig.~\ref{fig:ev}. The dashed line represents the
abundances of T cells targeting each viral phenotype, represented as vertical stems.}
  \label{fig:speciation}
\end{figure}

Finally we have studied how the co-evolutionary and competitive dynamics of viral strains, mediated by the immune response, may lead to the formation of new viral quasispecies.
In Fig.~\ref{fig:speciation} we consider a phenotypic space of 25 strains and the first inoculum is at phenotype 15 (Fig.~\ref{fig:speciation}a), followed by a delayed inoculum at phenotype~5 at time $t=1$.
We account for the differences in recognition ability of viral antigens by T cells by using a non-uniform interaction strength which favor the central phenotypes. The immune system does not discriminate among similar phenotypes, thus inducing a competition among neighboring strains. The result of this induced competition is the separation of the original quasispecies into two clusters (quasi-speciation), Fig.~\ref{fig:speciation}b. However the immune system response continues to change in time (Figs.~\ref{fig:speciation}c-d), resulting into a complex coevolution with viral populations.

\subsection{R5 to X4 switch}
\begin{figure}[t]
  \centering
  \includegraphics[width=0.8\textwidth]{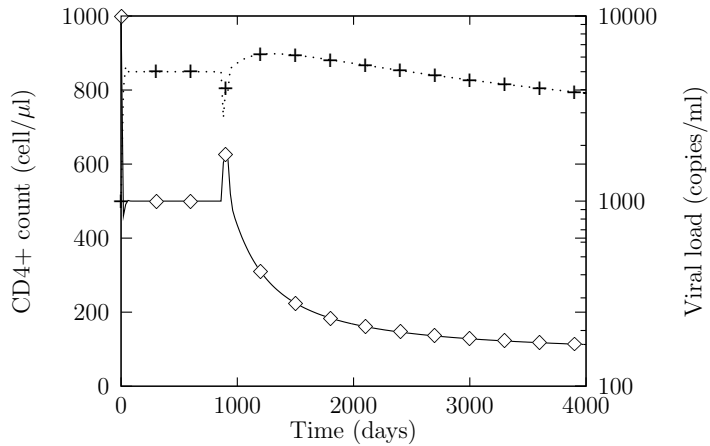}
  \caption{Time evolution of the concentrations of uninfected T-cells ($\Diamond$) and viruses (+), during R5 to X4 switch, occurring at time $t \approx 900$. The time of appearance of the X4 strains depends on the mutation rate and on the phenotypic distance between R5 and X4 viruses. After the  appearance of the X4 phenotype a continuous slow decline in CD4+ T-cells level leads to AIDS phase (CD4 counts below 200cells/ml). We set $\mu \approx 0.001$ and $d_P \approx 5$.}
  \label{fig:X4}
\end{figure}
We studied the coevolutive dynamics leading to X4 strain  appearance by successive mutations of the ancestor R5 strain. The stimulated production of TNF regulate the interactions between immune response and the virus and between the different strains of HIV virus. The results of these interactions are a decline in T-cells level, leading to the AIDS phase of the disease, and the decline in levels of viruses using the R5 coreceptor.
In Fig.~\ref{fig:X4} the temporal evolution of the infection is shown, with the  appearance of the X4 strain, and the successive decline in T-cells abundances.
The time at which the phenotypic switch occurs depends both on the mutation rate $\mu$ and on the phenotypic distance between R5 and X4 strains, $d_P$. Once experimental data are known, it's possible to tune the model parameters to their corresponding biological values. Moreover, to get a better insight on the range of variability of the time of appearance of X4 phenotype, a sensitivity analysis for varying values of the $\mu$ and $d_P$ parameters is still possible.

\begin{figure}[t]
  \centering
  \includegraphics[width=0.8\textwidth]{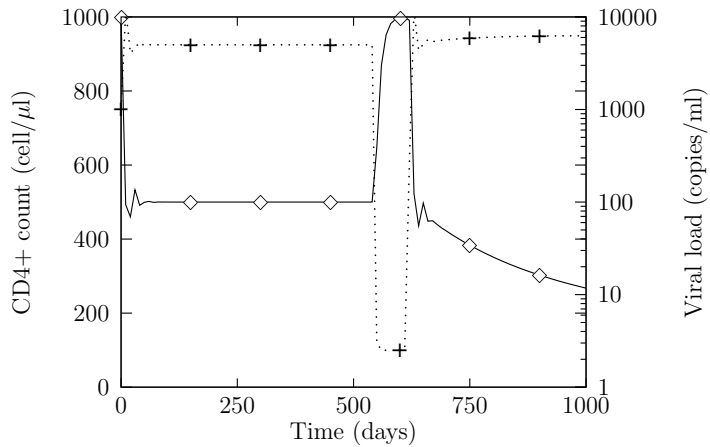}
  \caption{The efficacy of HAART therapy may be disrupted by a sudden interruption in drugs treatment. If time has passed for mutations to populate the R5 strains closer to the X4 phenotypes, an earlier appearance of X4 strains may occur. Uninfected T-cells ($\Diamond$) and viruses (+). Parameters as in Fig.~\ref{fig:X4}.}
  \label{fig:haart}
\end{figure}
In Fig.~\ref{fig:haart} we considered HAART treatment, which is usually able to decrease the concentration of the virus in the blood and delay the X4 appearance. The use of this model suggests a possible scenario in the case of a sudden interruption in the therapy. If the different R5 strains experience the same selection pressure, as soon as the therapy is stopped, the X4 strain may appear sooner. Infact during the treatment the concentration of the different strains of R5 viruses is kept to a very low level while T-cell abundances increase. As the therapy is interrupted, all the strains give rise to  a renewed infection, but now also the strains closer to the X4 co-receptor using viruses are populated, and a mutation leading to an X4 strain occurs sooner.

\begin{figure}[t]
  \centering
  \includegraphics[width=0.7\textwidth]{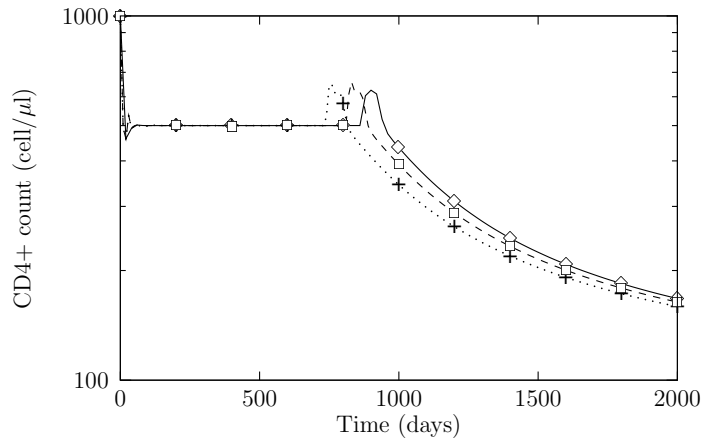}
  \caption{CD4+ T-cells concentration during HIV-1 super-infection by a R5 viral strain. Different signs represent: evolution without superinfection, ($\Diamond$); superinfection occurring at time $t$=100 and 400, (+) and ($\Box$), respectively. For a superinfection event occurring after the R5 to X4 switching the dynamics is qualitatively the same as for a single infection, ($\Diamond$). If the second delayed infection occurs before the R5 to X4 switching, the time of appearance of X4 viruses may be shorter, when the superinfecting strain is closer to the X4 phenotypes, (+,$\Box$). Parameters as in Fig.~\ref{fig:X4}.}
  \label{fig:superX4}
\end{figure}
We have finally studied the influence of switching co-receptor usage in superinfection dynamics. In Fig.~\ref{fig:superX4} we show T-cells dynamics for different times of the superinfection event. We may observe that if the superinfection occurs after the appearance of the X4, the new R5 strain does not have any effect on T-cells behavior. On the other hand is worth noting that if the new R5 inoculum take place before the X4 appearance, this may speed up the switching to the X4 phenotype if the new strain is mutationally close to the X4.

\subsection{Investigating the mutational pathway between R5 and X4}
Research into HIV dynamics has much to gain from investigating the
evolution of chemokine co-receptor usage. Although CCR5 and CXCR4
are the major coreceptors used by HIV-1 a number of chemokine
receptors display coreceptor activities in vitro. Several other
chemokine receptors, possibly not present on the T cell membrane,
may act as targets. To date, a number of human receptors, specific
for these chemokine subfamilies, have been described, though many
receptors are still unassigned. Several viruses, for example
Epstein-Barr, Cytomegalovirus, and Herpes Samiri, contain
functional homologous to human CRs, an indication that such
viruses may use these receptors to subvert the effects of host
chemokines~\cite{Mu2001}. Cells different from CD4 and CD8, such
as macrophages, express lower levels of CD4, CCR5, and CXCR4 on
the cell surface compared with CD4+ T cells~\cite{LC1996,OD1999,WG2002},
and low levels of these receptors
expressed on macaque macrophages can restrict infection of some
non-M-tropic R5 HIV-1 and X4 simian immunodeficiency virus (SIV)
strains~\cite{BS2000,MD2000}.
\begin{figure}
\centering
\includegraphics[width=0.8\textwidth]{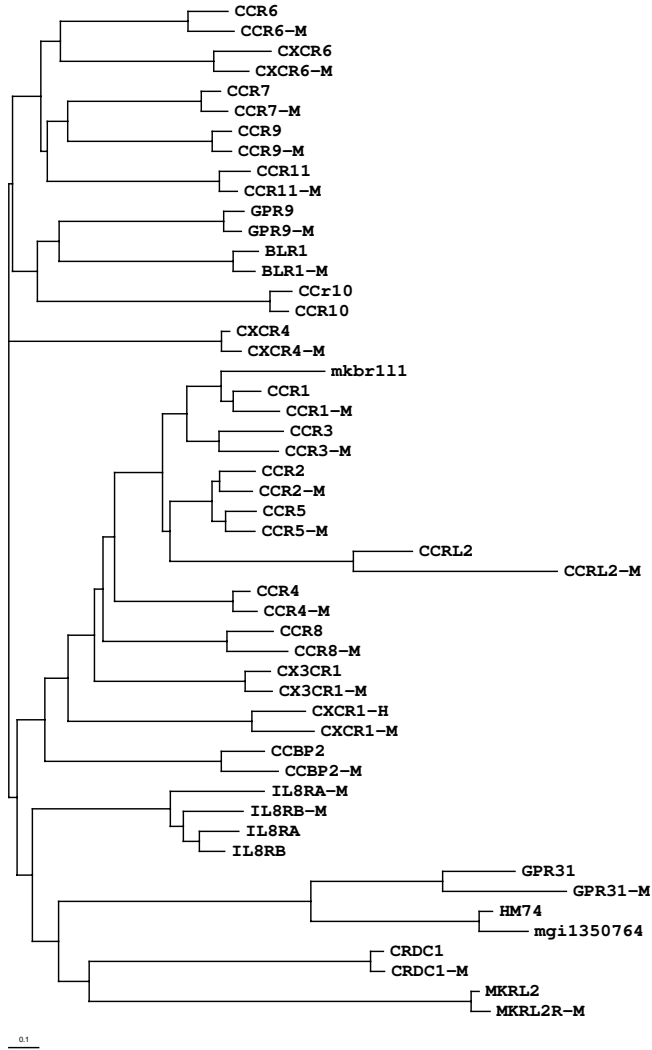}
\caption{The maximum likelihood phylogeny under the JTT+F+$\Gamma$
model of evolution for the set of human and mouse (mouse sequences
are labeled with "-mou") chemokine receptors amino acid sequences
from human and mouse. The scale bar refers to the branch lengths,
measured in expected numbers of amino acid replacements per site}
\label{fig:tree1}
\end{figure}
\begin{figure}
\centering
\includegraphics[width=0.7\textwidth]{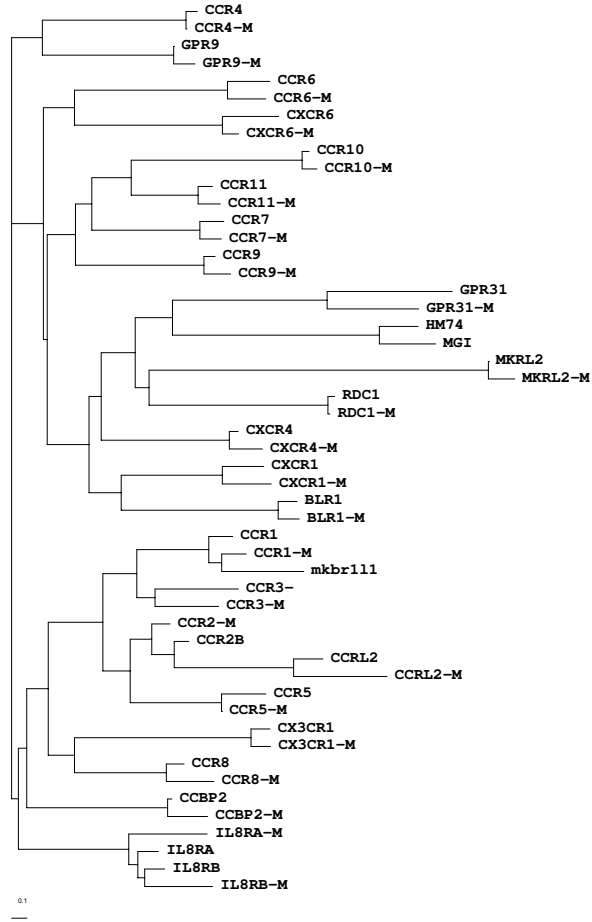}
\caption{The same as Fig.~\ref{fig:tree1} but considering only
the external loop regions.}
\label{fig:tree2}
\end{figure}

Fundamental to the evolutionary approach is the representation of
the evolution of sequences along lineages of evolutionary trees,
as these trees describe the complex patterns of dependence amongst
sequences that are caused by their common ancestry~\cite{WG2001,S2000,GC2000}.

The ML tree, obtained using the JTT+$F$+$\Gamma $ model of
evolution, is shown in Figure~\ref{fig:tree1}. The topology
clearly shows that the CCR family is not homogeneous: CCR6, CCR7,
CCR9 and CCR10 are separated from the other CCRs; in particular,
CCR10 clusters with CXCRs; CXCR4 and CXCR6 do not cluster with the
CXCRs. The tree shows that there are many mutational steps between
CCR5 and CXCR4. The phylogeny suggests that the mutations that
allow the virus env to cover a wide phenotypic distance from R5 to
X4, may also lead to visit other receptors. Since the external
loops of CRs contain the binding specificities and have higher
rates of evolution than internal loops and transmembrane segments
~\cite{SA1997}, the tree Fig.~\ref{fig:tree2} shows a relative
longer mutational pathway between CCR5 and CXCR4 with respect to
pathway linking CCR5 to other receptors.

\section{Discussion}
The worldwide presence of several strains of the HIV virus and
their often simultaneous presence within a patient, due to the
increased frequency of multiple infections, are the remarkable
features of HIV pandemia. For example, HIV-1 exists as several
groups, subdivided in growing number of subtypes which are
slightly predominant in different geographical regions
~\cite{YL2005}. HIV-1-infected CD4+ T cells isolated from the
spleens of two individuals were recently shown to harbor anywhere
between one and eight proviruses, with an average of three to four
proviruses per cell ~\cite{DP2005}. Mutations, recombinations and
selection pressure cause the appearance of
quasispecies~\cite{ES1989}.

The interest in HIV quasispecies is motivated by concern about
developing strain specific drugs. Quasispecies are likely the key
for understanding the emerging infectious diseases and has
implications for transmission, public health counseling,
treatment and vaccine development. Moreover, the observed
co-evolutionary dynamics of virus and immune response opens the
way to the challenging possibility of the introduction or
modulation of a quasispecies to be used in therapy against an
already present aggressive strain, as experimented by Snell and
colleagues~\cite{SR1997}. The authors showed that the introduction
of an engineered virus can achieve HIV load reduction of 92\% and
recovery of host cells to 17\% of their normal levels (see also
the mathematical model in Ref.~\cite{RG2003}).

Different drug treatments can alter the population of
quasispecies. Will R5 blocking drugs cause HIV to start using X4?
And will that be worse than letting the R5-using virus stay around
along at its own, slower, but no less dangerous activity?

We first presented a model of the within-patience persistence of HIV quasispecies, by extending to multiple strain the Perelson's standard model~\cite{PH1996}. This approach allows to incorporate coevolutive and competitive dynamics resulting from the different strains of HIV virus and different antigen recognition abilities by the immune system considered. Our model shows that the time evolution of the competition between quasispecies is slow and has time scales of several months.

Recent works show that TNF is a prognostic marker for the
progression of HIV disease~\cite{HS2005a,HS2005b}.  We focused on
both the inability of the thymus to efficiently compensate for
even a relatively small loss of T cells precursors and on the role
of TNF in regulating the interactions between the different
strains of HIV virus. The second model we have introduced shows
that keeping low the concentration of TNF, both the depletion of
T-cells precursors repertoire and the R5 overcome by X4 strains
slow down.

Phylogenetic inference of chemokine receptors shows that there are
several mutational patterns linking CCR5 to several receptors that
have the same branch length of that from CCR5 to CXCR4. There is a
massive abundance of signaling disruptions in the immune systems
during AIDS progression, particularly after the transition R5 to
X4. These disruptions may be due to variants of the virus which
bind other chemokine receptors. This hypothesis also suggests that
R5-late strains in not-X4 AIDS, which are known to be different
from R5 early strains, may have accumulated mutations enabling them
to interact with other chemokine receptors. Therefore, our model
suggests the sooner the HAART the better, because the presence of
a large number of R5 will increase the mutational spectra in R5
strains (late R5) and the probability of getting closer to the
binding specificities of other chemokine receptors.

The large effect of TNF on T cells dynamics described by our
model, suggests the benefit of a TNF buffering therapy. It is
known that the dynamics of TNF is related to the dynamics of
TNF-related apoptosis-inducing ligand (TRAIL). In this model we
consider constant the concentration of TRAIL~\cite{HS2005b} and we
do not consider many other important players such as Rantes.

Our models represent also a general framework to investigate
intermittency or switching dominance of strains and the arising of
new dominant strains during different phases of therapy; how
superinfection will evolve in case of replacement of
drug-resistant virus with a drug-sensitive virus and acquisition
of highly divergent viruses of different strains; to investigate
whether antiviral treatment may increase susceptibility to
superinfection by decreasing antigen load.

Work in progress focuses on refining the "R5toX4" model by
incorporating the dynamics for TRAIL and comparing our results
with those of Ribeiro and colleagues~\cite{RD2006} who have
presented a model of R5 to X4 switch based on the hypothesis that
X4 and R5 viruses have a preferential tropism for naïve and memory
T cells, respectively.

\section{Acknowledgements}
We thank Stefano Ruffo for insightful suggestions.


\newpage

\begin{thebibliography}{00}

\bibitem{AW2002}  Altfeld, M., T. M. Allen, X. G. Yu, M. N. Johnston, D. Agrawal, B. T. Korber, D. C. Montefiori, D. H. O'Connor, B. T. Davis, P. K. Lee, E. Maier, J. Harlow, P. J.R. Goulder, C. Brander, E. S. B. Rosenberg D. and. Walker. 2001. Evolution and transmission of stable CTL escape mutations in HIV infection. Nature \textbf{420}:434-438.

\bibitem{Ba1998} Baggiolini, M. 1998. Chemokines and leukocyte traffic. Nature \textbf{392}:565-568.

\bibitem{BB1997} Bagnoli, F., and M. Bezzi. 1997. Speciation as Pattern Formation by Competition in a Smooth Fitness Landscape. Phys. Rev. Lett. \textbf{79}:3302-3305.

\bibitem{BL2005}  Bello, G., C. Casado, V. Sandonis, M. Alonso-Nieto, J. L. Vicario, S. Garc\'{i}a, V. Hernando, C. Rodr\'{i}guez, J. R. delomero C. and L\'{o}pez-Gal\'{i}ndez. 2005. A subset of human immunodeficiency virus type 1 long-term non-progressors is characterized by the unique presence of ancestral sequences in the viral population. J. Gen. Virol. \textbf{86}:355-364.

\bibitem{BS2005} Bagnoli, F., P. Lio' and L. Sguanci. 2005. Modeling viral coevolution: HIV multi-clonal persistence and competition dynamics. Phys. A, In Press, Corrected Proof, Available online 28 November 2005.

\bibitem{BS2000} Bannert, N., D. S. Schenten, J. Craig and J. Sodroski. 2000. The level of CD4 expression limits infection of primary rhesus monkey macrophages by a T-tropic simian immunodeficiency virus and macrophagetropic human immunodeficiency viruses. J. Virol. \textbf{74}:10984-10993.

\bibitem{BE2005} Biebricher CK, M. Eigen .2005. The error threshold. Virus Res. 107:117-27.

\bibitem{BC2000} Burch, C. L., and L. Chao. 2000. Evolvability of an RNA virus determined by its mutational neighborhood. Nature \textbf{406}:625-628.

\bibitem{CS1996} Celada, F. , and P. E. Seiden. 1996. Affinity maturation and hypermutation in a simulation of the humoral immune response. Eur. J. Immunol. \textbf{26}:1350-1358.

\bibitem{CP2004}  Chao, L., M. P. Davenport, S. Forrest, A. S. and. Perelson. 2004. A stochastic model of cytotoxic T
cell responses. J Theor Biol. 228:227-240.

\bibitem{CD2000}  Costa, L., M. Janini, M. Pinto R. and Diaz. 2000. Quasispecies analyses of a familial cluster infected with multiple HIV-1 subtypes. Int. Conf. AIDS. \textbf{13}:(abstract no. MoPpA1005).

\bibitem{DO1978} Dayhoff, M. O., R. M. Schwartz and B. C. Orcutt. 1978. A model of evolutionary change in
proteins. In Atlas of Protein Sequence and Structure, (Vol. 5, Suppl. 3) (Dayhoff, M.O., ed.), pp. 345.352, National Biomedical Research Foundation.

\bibitem{DP2005} Dixit, N. M., and A. S. Perelson. 2005. HIV dynamics with multiple infections of
target cells. Proc. Natl. Acad. Sci. USA. 102:8198-203.

\bibitem{DP1992}  DeBoer, R. J., L.A. Segel and A. S. Perelson, J. 1992. Pattern formation in one and two-dimensional shape-space models of the immune system. J. Theor. Biol. 155:295-333.

\bibitem{DP1995}  DeBoer, R. J., and A. S. Perelson, J. 1995. Towards a general function describing T-cell proliferation. J. Theor. Biol. 175:657-576.

\bibitem{ES1977} Eigen, M., and  P. Schuster. 1977. The hypercycle. Naturwissenschaften \textbf{64}:541-565.

\bibitem{ES1989} Eigen, M., J. McCaskill and P. Schuster. 1989. The Molecular Quasi-Species. Adv. Chem. Phys. 75:149-263.

\bibitem{FP1997} Franz, S., and  L. Peliti. 1997. Error Threshold in Simple Landscapes. J. Phys. A \textbf{30}:4481-4487.

\bibitem{GH2005} Gerhardt, M., D. Mloka, S. Tovanabutra, E. Sanders-Buell, O. Hoffmann, L. Maboko, D. Mmbando, D. L. Birx, F. E. McCutchan and M. Hoelscher. 2005. In-depth, longitudinal analysis of viral quasispecies from an individual triply infected with late-stage human immunodeficiency virus type 1, using a multiple PCR primer approach. J. Virol.\textbf{79}:8249-8261.

\bibitem{GG2002} Gog, J. R., and B. T. Grenfell. 2002. Dynamics and selection of many-strain pathogens. Proc. Nat. Acad. Sci. \textbf{99}:17209-17214.

\bibitem{GC2000} Goh, C. S., A. A. Bogan, M. Joachimiak, D. Walther, F. E. Cohen. 2000.  Co-evolution of proteins with their interaction partners. \textit{J. Mol. Bio.} \textbf{299}:283-293.

\bibitem{GG2005} Gorry, P.R., M. Churchill, S. M. Crowe, A. L. Cunningham and D. Gabuzda. 2005. Pathogenesis of macrophage tropic HIV. Curr. HIV Res. 3:53-60.

\bibitem{GS2005} Gray, L., J. Sterjovski, M. Churchill, P. Ellery, N. Nasr, S. R. Lewin, S. M. Crowe, S. L. Wesselingh, A. L. Cunningham and P. R. Gorry. 2005. Uncoupling coreceptor usage of human immunodeficiency virus type 1 (HIV-1) from macrophage tropism reveals biological properties of CCR5-restricted HIV-1 isolates from patients with acquired immunodeficiency syndrome. Virology 337:384-98

\bibitem{HH2004} Haaft, P. T., E. J. Verschoor, B. Verstrepen, H. Niphuis, R. Dubbes, W. Koornstra, W. Bogers, B. Rosenwirth  and J. L. Heeney. 2004. Readily acquired secondary infections of human and simian immunodeficiency viruses following single intravenous exposure in non-human primates. J. of Gen. Virol. \textbf{85}:3735-3745.

\bibitem{HS2005a} Herbeuval, J. P., A. W. Hardy, A. Boasso, S. A. Anderson, M. J. Dolan, M. Dy, and G. M. Shearer. 2005. Regulation of TNF-related apoptosis-inducing ligand on primary CD4+ T cells by HIV-1: Role of type I IFN-producing plasmacytoid dendritic cells. Proc. Nat. Acad. Sci. USA \textbf{102}:13974-13979.

\bibitem{HS2005b} Herbeuval, J. P., Grivel, J.-C., Boasso A., Hardy, A. W. , Chougnet, C., Dolan M. J., Yagita, H., Lifson J. D., and
Shearer, G. M. 2005. CD4+ T-cell death induced by infectious and
noninfectious HIV-1: role of type 1 interferon-dependent,
TRAIL/DR5-mediated apoptosis. Blood 106: 3524-3531.

\bibitem{HM1995}  Ho, D., A. U. Neumann, A. S. Perelson, W. Chen, J. M. Leonard M. and Markowitz. 1995. Rapid turnover of plasma virions and CD4 lymphocytes in HIV-1 infection. Nature \textbf{373}:123-126.

\bibitem{JD2000}  Jetzt, E., H. Yu, G. J. Klarmann, Y. Ron, B. D. Preston,  J. P. and. Dougherty. 2000. High Rate of Recombination throughout the Human Immunodeficiency Virus Type 1 Genome. J. Virol. \textbf{74}:1234-1240.

\bibitem{JT1994} Jones DT, WR Taylor, JM Thornton .1994. A mutation data matrix for
transmembrane proteins. FEBS Lett. 339:269-75.


\bibitem{KF2004} Karlsson, I., L. Antonsson, Y. Shi, M. Oberg, A. Karlsson, J. Albert, B. Olde, C. Owman, M. Jansson and E. M. Fenyo. 2004. Coevolution of RANTES sensitivity and mode of CCR5 receptor use by human immunodeficiency virus type 1 of the R5 phenotype. J. Virol. 78:11807-11815.

\bibitem{KH1997} Krempl, C., B. Schultze, H. Laude and G. Herrler. 1997. Point mutations in the S protein connect the sialic acid binding activity with the enteropathogenicity of transmissible gastroenteritis coronavirus. J. Virol. \textbf{71}:3285-3287.

\bibitem{L2003} Levy, J. A. 2003. Is HIV superinfection worrisome? The Lancet \textbf{361}:98-99.

\bibitem{LC1996} Lewin, S.R., S. Sonza, L. B. Irving, C. F. McDonald, J. Mills and S. M. Crowe. 1996. Surface CD4 is critical to in vitro HIV infection of human alveolar macrophages. AIDS Res. Hum. Retroviruses 12:877-883.

\bibitem{LL2002}  Lin, J., V. Andreasen,R. Casagrandi L. and S.A. Levin. 2002. Traveling waves in a model of influenza A drift. J. Theor. Biol. \textbf{222}:437-445.

\bibitem{Luster1998} Luster, A. D. 1998. Chemokines: chemotactic cytokines that mediate inflammation. N. Engl. J. Med. \textbf{338}:436-445.

\bibitem{Mellado2001} Mellado, M., Rodriguez-Frade J. M., Vila-Coro A. J., S. Fernandez, A. Martin de Ana, D. R. Jones, J. L. Toran, A.C. Martinez .2001. Chemokine receptor homo- or heterodimerization activates distinct signaling pathways. \textit{J. EMBO} \textbf{20}:2497-507.

\bibitem{MD2000} Mori, K., M. Rosenzweig and R.C. Desrosiers. 2000. Mechanisms for adaptation of simian immunodeficiency virus to replication in alveolar macrophages. J. Virol. \textbf{74}:10852-10859.

\bibitem{Mu2001} Murphy, P. M. 2001. Viral exploitation and subversion of the immune system through chemokine mimicry. Nature Immunology 2:116-122.

\bibitem{NP1998} Neumann, A. U., N. P. Lam, H. Dahari, D. R. Gretch, T. E. Wiley, T. J. Layden,  and A. S. Perelson. 1998. Hepatitis C Viral Dynamics in Vivo and the Antiviral Efficacy of Interferon-alpha Therapy. Science \textbf{282}:103-107.

\bibitem{OD1999} Ometto, L., M. Zanchetta, A. Cabrelle, G. Esposito, M. Mainardi,
L. Chieco-Bianchi and A. De Rossi. 1999. Restriction of HIV type 1 infection in macrophages heterozygous for a deletion in the CCchemokine receptor 5 gene. AIDS Res. Hum. Retroviruses 15:1441-1452.

\bibitem{PY2004} Pavlakis, G. N., A. Valentin, M. Morrow and R. Yarchoan.  Differential effects of TNF on HIV-1 expression: R5 inhibition and implications for viral evolution. Int Conf AIDS 2004 Jul 11-16;
15:(abstract no. MoOrA1048).

\bibitem{PH1996} Perelson, S., A. U. Neumann, M. Markowitz, J. M. D. Leonard D. and. Ho. 1996. HIV-1 Dynamics in Vivo: Virion Clearance Rate, Infected Cell Life-Span, and Viral Generation Time. Science \textbf{271}:1582-1586.

\bibitem{R2002}  Ramos, A., D. J. Hu, L. Nguyen,K. O. Phan, S. Vanichseni, N. Promadej, K. Choopanya, M. Callahan, N. L. Young, J. McNicholl, T. D. Mastro, T. M. Folks and S. Subbarao. 2002. Intersubtype human immunodeficiency virus type 1 superinfection following seroconversion to primary infection in two injection drug users. J. Virol. \textbf{76}:7444-7452.

\bibitem{RG2003} Revilla, T. , and  G. Garc\'{y}a-Ramos. 2003. Math. Biosc. \textbf{185}:191.

\bibitem{RD2006} Ribeiro, M. R., D. M. Hazenberg, S. A. Perelson, and  P. M. Davenport.  2006. Naïve and Memory Cell Turnover as Drivers of CCR5-to-CXCR4 Tropism Switch in Human Immunodeficiency Virus Type 1: Implications for Therapy.  J. Virol. \textbf{80}:802-809.

\bibitem{SA1997} Samson, M., G. LaRosa, F. Libert, P. Paindavoine, G. Detheux, R. Vassay, M.  Pandarmentier. 1997. The second extracellular loop of CCR5 is the major determinant of ligand specificity. \textit{J. Biol. Chem} \textbf{272}:24934-24941.

\bibitem{SR1997}  Schnell, J., E. Johnson, L. Buonocore and  J. K. Rose. 1997. Construction of a novel virus that targets HIV-1-infected cells and controls HIV-1 infection. Cell \textbf{90}:849-857.

\bibitem{S2000} Shields, D.C. 2000. Gene conversions among chemokine genes. Gene \textbf{246}:239-245.

\bibitem{SP1997} Speck, R. F., K. Wehrly, E. J. Platt, R. E. Atchison, I. F. Charo, D. Kabat, B. Chesebro, M. A. Goldsmith. 1997. Selective employment of chemokine receptors as human immunodeficiency virus type 1 coreceptors determined by individual amino acids within the V. envelope3 loop. J. Virol. \textbf{71}:7136-7139.

\bibitem{WG2002} Wang, J., K. Crawford, M. Yuan, H. Wang, P. R. Gorry and D. Gabuzda. 2002. Regulation of CC chemokine receptor 5 and CD4 expression and human immunodeficiency virus type 1 replication in human macrophages and microglia by T helper type 2 cytokines. J. Infect. Dis. 185:885-897.

\bibitem{WS1995}  Wei, X., S. K. Ghosh, M. E. Taylor, V. A. Johnson, E. A. Emini, P. Deutsch, J. D. Lifson, S. Bonhoeffer, M. A. Nowak, B. H. Hahn, M. S. Saag and G.M. Shaw. 1995. Viral dynamics in human immunodeficiency virus type 1 infection. Nature \textbf{373}:117-122.

\bibitem{WG2001} Whelan, S. and N. Goldman. 2001. A general empirical model of protein evolution derived from multiple protein families using a maximum likelihood approach. Mol Biol Evol. \textbf{18}:691-699.

\bibitem{WLG2001}  Whelan, S., P. Li\`{o}, N. Goldman. 2001. P. Molecularhylogenetics: state-of-the-art methods for looking into the past. \textit{G. Trendsenet.} \textbf{17}:262-272.

\bibitem{WP2004}  Wiegel, F. W., and A. S. Perelson. 2004. Some scaling principles for the immune system. Immunol. Cell. Biol. \textbf{82}:127-131.

\bibitem{WN2002}  Wodarz, D., and M. A. Nowak. 2002.  HIV dynamics and evolution. BioEssays \textbf{24}:1178-1187.

\bibitem{Y1994}  Yang, Z. (1994) Maximum likelihood phylogenetic estimation from DNA sequences with variable rates over sites: approximate methods. J. Mol. Evol. 39:306-314

\bibitem{Y1997} Yang, Z. (1997) PAML: a program package for phylogenetic analysis
by maximum likelihood. CABIOS 13:555-556

\bibitem{YL2005} Yang, O.O., E. S. Daar, B. D. Jamieson, A. Balamurugan, D. M. Smith, J. A. Pitt, C. J. Petropoulos, D. D. Richman, S. J. Little and A. J. Leigh-Brown. 2005. Human immunodeficiency virus type 1 clade B superinfection: evidence for differential immune containment of distinct clade B strains. J.  Virol. \textbf{79}:860-868.

\bibitem{VA2006} Vignuzzi, M., J. K. Stone, J. J. Arnold, C. E. Cameron and R. Andino. 2006. Quasispecies diversity determines pathogenesis through cooperative interactions in a viral population. Nature 439:344-348.

\end{thebibliography}
\end{document}